\documentclass[twocolumn,showpacs,preprintnumbers,amsmath,amssymb]{revtex4}

\usepackage{graphicx}
\usepackage{dcolumn}
\usepackage{bm}

\begin{document}


\title{Universality in Dynamical Formation of Entanglement for Quantum Chaos
}

\author{Hiroto Kubotani}
\email{kubotani@yukawa.kyoto-u.ac.jp}
\affiliation{%
Institute of Physics, Faculty of Engineering, Kanagawa University, \\
Yokohama 221-8686, Japan
}%

\author{Mikito Toda}%
\email{toda@ki-rin.phys.nara-wu.ac.jp}
\affiliation{%
Department of Physics, Faculty of Science, Nara Womens University,
Nara 630-8506, Japan
}%

\author{Satoshi Adachi}
\email{adachi@aa.ap.titech.ac.jp}
\affiliation{
Department of Physics, Faculty of Science, Tokyo Institute of Technology,  
Meguro 152-8550, Japan
}%

\date{\today}

\begin{abstract}
Dynamical formation of entanglement is studied
for quantum chaotic bi-particle systems.
We find that statistical properties of the Schmidt eigenvalues
for strong chaos are well described 
by the random matrix theory of the Laguerre ensemble.
This implies that entanglement formation for quantum chaos 
has universal properties,
and does not depend on specific aspects of the systems.
\end{abstract}

\maketitle


At the dawn of the quantum mechanics, 
entangled states of spatially separated bi-particle systems triggered
fierce discussions because of their paradoxical property \cite{EPR}. 
These discussions lead to the finding of the inequalities
which measure how quantum systems deviate 
from the classical reality \cite{Bell}.
Recently, a renewed interest has been paid to entanglement 
since it is recognized as an invaluable resource 
in quantum computation and quantum communication \cite{Nielsen}.
Thus, the information theory and the circuit theory 
for entangled qubits has become fast growing fields.

On the other hand, in the quantum theory of open systems,
dynamical formation of entanglement between a system and its environment 
is a target of the research.
In these studies, it is supposed that uncontrolled formation of entanglement
causes the loss of quantum coherence in the system 
and leads to the appearance of the classical reality.
Then, the loss of quantum coherence is considered to provide
the foundation of the statistical physics \cite{Zeh,Zurek}.
Also in the quantum computation, 
dynamical formation of entanglement between qubits and their environment 
is a crucial subject
since it can give rise to erroneous operations of quantum computers.

In the traditional study on the dynamical formation of entanglement,
the system is supposed to interact with an environment
composed of a large number of degrees of freedom.
Then, the interaction with the environment causes
complicated uncontrollable formation of entanglement,
leading to the loss of coherence in the system. 

However, interaction with an environment is not the only cause of
complicated processes of forming entanglement.
Actually, dynamical formation of entanglement 
by quantum chaos has been an active topic these days 
\cite{Adachi92,Kubotani95,Furuya98}.
In their studies, quantum chaos 
gives rise to increase of the von Neumann entropy 
(or a modified version of it) 
to the extent that the state approaches the maximally entangled state, 
which is an equally weighted superposition of entangled bases.
Therefore, quantum chaotic systems can also be the origin of the loss of 
coherence, 
even when they are of a small number of degrees of freedom.

This indicates that there are several causes
where complicated processes of forming entanglement takes place.
Then, the next question is how we can classify, if possible, 
those processes of forming complicated entanglement.
Are there any universality classes according to which
we can differentiate seemingly similar processes of forming entanglement ?

In this letter, we show that there actually exists
a universal class in the dynamical processes of forming entangled states.
Here, we investigate the time evolution of Schmidt eigenvalues
for coupled systems of quantum chaos,
and indicate that, for strongly chaotic systems,
the distribution of Schmidt eigenvalues approaches 
a class of random matrices, i.e., the Laguerre unitary ensemble 
\cite{Bronk65}. 
So far, the Laguerre unitary ensemble has received only a little attention
in the theory of random matrices \cite{Haake01,Mehta91}.
We will discuss possible implications of this universal class,
and suggest its extensions for other cases.

We consider a wave function of a bi-particle system, $\Phi(x_1, x_2)$. 
For simplicity, we assume that the dimensions of the Hilbert spaces
for each degree of freedom are the same, 
and will be denoted by $N$. 
Then, the wave function is represented as 
\begin{equation}
\Phi(x_1,x_2)=\sum_{i,j}^{N}A_{ij}|i>_{1}\otimes|j>_{2} .
\end{equation}
Here $\{|i>_1|i=1, 2,\ldots, N\}$ and $\{|j>_2|j=1, 2,\ldots, N\}$ 
are the orthogonal bases for each degree of freedom, and 
$A_{ij} (i, j=1, 2,\ldots, N)$  are complex-valued coefficients.
For the $N\times N$ matrix $A$ whose components are $A_{ij}$, 
we perform the singular value decomposition \cite{MatrixComputations},
\begin{equation}
A =U \Lambda V^{-1}
\label{SVD}
\end{equation}
where $U$ and $V$ are $N\times N$ unitary matrices, and 
$\Lambda$ is a diagonal one whose components are $\lambda_i (i=1, 2,\ldots, N)$. 
Here, we can always make $\lambda_i$ 
non-negative real and 
set $\lambda_i \ge \lambda_j$ for $i <j$.
We transform the bases from $\{|i>_1\}$ and $\{|j>_2\}$ to 
$\{|\tilde{i}>_1\}$ and $\{|\tilde{j}>_2\}$ using $U$ and $V$, respectively 
so that the wave function is described as the Schmidt decomposition 
form,
\begin{equation}
\Phi(x_1,x_2)=\sum_{i}^{N}\lambda_i|\tilde{i}>_{1}\otimes|\tilde{i}>_{2} .
\label{SchmidtDecomp}
\end{equation}
For the representation (\ref{SchmidtDecomp}), 
the von Neumann entropy for the reduced density matrix, in which 
one of the degree of freedom, $x_1$ or $x_2$ is traced out, 
is easily evaluated as 
\begin{equation}
S=-\sum_i^N \lambda_i^2 \ln(\lambda_i^2).
\label{vonNeumannEntropy}
\end{equation}
We note that only the diagonal matrix $\Lambda$ contributes to $S$.


From now on, 
we construct a random matrix theory for the coefficient matrices, 
$A$ of wave functions.
First, we assume that the occurrence probability for 
the matrix $A$ is a function of the independent components $A_{ij}$.
Moreover, we assume that the probability is 
invariant under local unitary transformations.
These assumptions determine the joint probability function $P(\{A_{ij}\})$ 
as 
\begin{equation}
P(\{A_{ij}\})\prod_{i,j}dA_{ij}
=C \exp [ -{1\over 2\alpha^2}{\rm Tr}(A^{\dagger}A)]
\prod_{i,j}dA_{ij}, 
\end{equation}
where $\alpha$ and $C$ are constants.
Transforming the arguments of $P$ from 
$A_{ij}$ into $\lambda_i$ and the parameters which determine $U$ and $V$,
and integrating the probability with respect to 
the variables other than $\lambda_i$, 
we obtain the probability function with respect to 
the eigenvalues $\varepsilon_i$,
\begin{equation}
P_L(\{\varepsilon_{i}\})\prod_{i}d\varepsilon_{i}\\
\equiv C' \prod_{i<j}|\varepsilon_i-\varepsilon_j|^2 e^{-\sum_i \varepsilon_i}
\prod_i d\varepsilon_i,
\label{LaguerreDist}
\end{equation}
where we introduced the new variables $\varepsilon_i$($\equiv N^2\lambda_i^2$), 
which take the values from 0 to $\infty$, and 
the normalization constant $C'$ \cite{Bronk65}\cite{Kubotani05}.
From the joint probability (\ref{LaguerreDist}), 
we can get the $n$-point correlation function as 
\begin{equation}
\begin{split}
&R_n(\varepsilon_1,\ldots, \varepsilon_n)\\
&\equiv {N! \over (N-n)!}\int_0^\infty \cdots \int_0^\infty
P_L(\varepsilon_1,\ldots, \varepsilon_N) d\varepsilon_{n+1}\cdots d\varepsilon_N\\
&={\rm det}[ K_N(\varepsilon_i, \varepsilon_j) ]_{i, j=1,\ldots, n} ~,
\end{split}
\label{npoint}
\end{equation}
where
\begin{equation}
K_N(\varepsilon_i, \varepsilon_j)\equiv \sum_{k=0}^{N-1}\varphi_k(\varepsilon_i)\varphi_k(\varepsilon_j)~.
\end{equation}
\label{Kdef}
Here, we take 
$\varphi_k(\varepsilon)\equiv e^{-\varepsilon/2}\mathcal{L}_k(\varepsilon)$
with $\mathcal{L}_k(\varepsilon)$ being the $k$-th normalized 
Laguerre polynomial,
hence we call the distribution (\ref{LaguerreDist}) 
the Laguerre unitary ensemble.
From Eq.~(\ref{npoint}), 
we obtain the one-level density \cite{Adachi05}
and 
the two-level cluster function \cite{Mehta91} as
\begin{equation}
\sigma_N(\varepsilon)\equiv R_1(\varepsilon)=K_N(\varepsilon,\varepsilon)
\label{LaguerreSigma}
\end{equation}
and 
\begin{equation}
T_2(\varepsilon_1, \varepsilon_2)
\equiv-R_2(\varepsilon_1,\varepsilon_2)+R_1(\varepsilon_1)R_1(\varepsilon_2)
=[K_N(\varepsilon_1, \varepsilon_2)]^2, 
\end{equation} respectively.
In order to discuss the statistical property of the level spacing, 
we need to unfold the level spectrum \cite{Haake01}.
Thus we introduce the rescaled level $\omega(\varepsilon)$ as follows
\begin{equation}
\omega(\varepsilon)=\int_0^\varepsilon\sigma_N(\varepsilon')d\varepsilon'.
\end{equation}
According to Nagao and Slevin \cite{Nagao93}, we also introduce 
the renormalized two-level cluster function as 
\begin{equation}
\bar{T}_2(\omega,\omega')
={T_2(\varepsilon(\omega),\varepsilon(\omega'))
\over \sqrt{R_1(\varepsilon(\omega))R_1(\varepsilon(\omega'))}}.
\label{RenormaizedClusterfunction}
\end{equation}



In comparing the Laguerre unitary ensemble with an 
ensemble of wavefunctions,
we will divide the distribution of the Schmidt eigenvalues 
into the following three regions:
The hard edge, i.e., those eigenvalues lying near zero,
the soft edge, i.e., those eigenvalues lying near the largest one,
and the bulk region, i.e., those lying 
away from the hard and soft edges.
Note that specific features of the Laguerre unitary ensemble 
emerge only in the region where eigenvalues are near zero, i.e.,
the hard edge.
For the other regions, it is known that 
characteristics of the Laguerre unitary ensemble 
are similar to those of the Gaussian unitary ensemble \cite{Nagao93}.

In the following, we study the dynamical formation of entanglement
for a coupled system of two degrees of freedom,
and compare the Schmidt eigenvalues of wavefunctions
to the Laguerre unitary ensemble .
We give the Hamiltonian as
\begin{equation}
H_{T}=H_1(x_1,p_1)+H_2(x_2,p_2)+H_{12}(p_1,p_2), 
\label{Htot}
\end{equation}
where
\begin{gather}
H_i(x_i,p_i)=2\sin^2({p_i\over 2})
+k_i\sum_{n=1}^\infty\sin(x_i)\delta(t-n)~~(i=1, 2), \label{Hkick}\\
H_{12}(p_1,p_2)=4c_{pp}\sin({p_1\over 2})\sin({p_2\over 2}).\label{Hint}
\end{gather}
Here $k_i~(i=1, 2)$ are the parameters for the kick strength and 
$c_{pp}$ is a coupling constant.
In order to make the dimension of the Hilbert space finite $N$,
we impose the periodic boundary conditions 
not only for $x_i$ but also for $p_i$, i.e.,
$0\le x_i < 2\pi$ and $-\pi\le p_i <\pi$~($i=1,2$).
These conditions require that the Planck constant be $\hbar =2\pi/N$, 
and we choose $N=128$.
For comparison 
between strong chaos 
and weak chaos, we choose the parameters as follows.
For strong chaos, their values are $k_1=3.0$, $k_2=2.5$ 
and, for weak chaos, they are $k_1=0.7$, $k_2=0.2$.
In both cases, we choose $c_{pp}=0.05$. 

\begin{figure}
\includegraphics[height=5.0cm,clip]{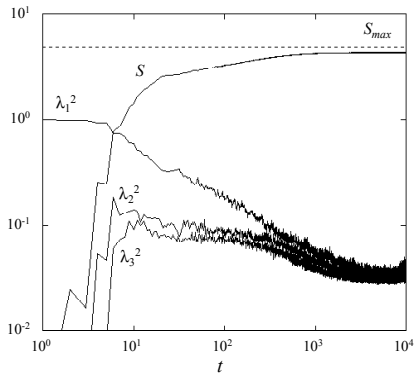}
\caption{\label{fig:S} Time evolution of $S$ and $\lambda_i^2$. 
The time evolution of $S$, $\lambda_1^2$, $\lambda_2^2$ and $\lambda_3^2$ 
is plotted for the strong chaos with $N=128$. 
The initial state is a product of the coherent states centered at 
$(x_i, p_i)=(\pi/2,\pi/4)~(i=1,2)$.
The increment of $S$ is slowly saturated before reaching the maximal entropy 
$S_{max}=\ln(N)$.
}
\end{figure}

As seen in Fig.~\ref{fig:S}, 
the time evolution of (\ref{vonNeumannEntropy}) by the strong chaos
shows a temporal generation of entanglement between $x_1$ and $x_2$, 
where we choose as an initial state 
a product of two coherent states. 
We notice that 
the variation of the entropy consists of two stages.
While the entropy increases rapidly in the initial stage, 
it starts to saturate for the latter stage.
In the initial stage, the largest Schmidt eigenvalue decreases
rapidly with the other ones growing from zero.
For the latter stage, these Schmidt eigenvalues start 
to avoid each other \cite{Kubotani05}.
The same behavior was found in a similar system \cite{Adachi05}.
Such avoided crossings remind us of the behavior of the energy eigenvalues
for quantum chaotic systems.
For weak chaos, saturation in the increase of entropy
also takes place.

Both for weak chaos and strong chaos, 
we construct an ensemble of $10^6$ wavefunctions, respectively, 
by collecting them after the entropy starts to saturate.
From the ensembles of the numerical data, 
we evaluate the one-level density 
$R_1(\varepsilon)$ and the two-level correlation function 
$R_2(\varepsilon, \varepsilon')$ directly, and estimate the renormalized 
two-level cluster function $\bar{T}_2(\omega,\omega')$ through 
Eq.~(\ref{RenormaizedClusterfunction}).

\begin{figure*}
\includegraphics[height=5.0cm,clip]{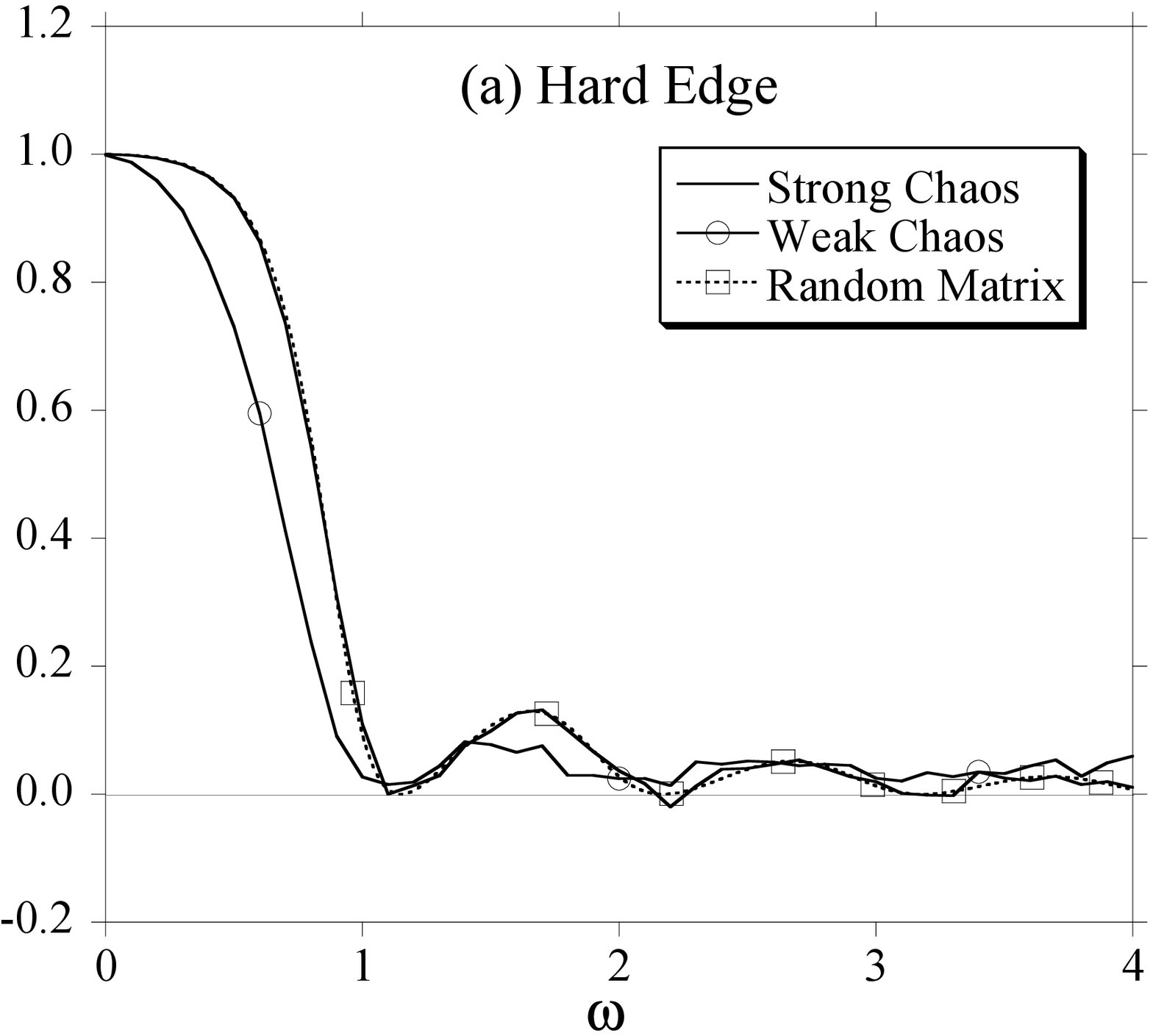}
\includegraphics[height=5.0cm,clip]{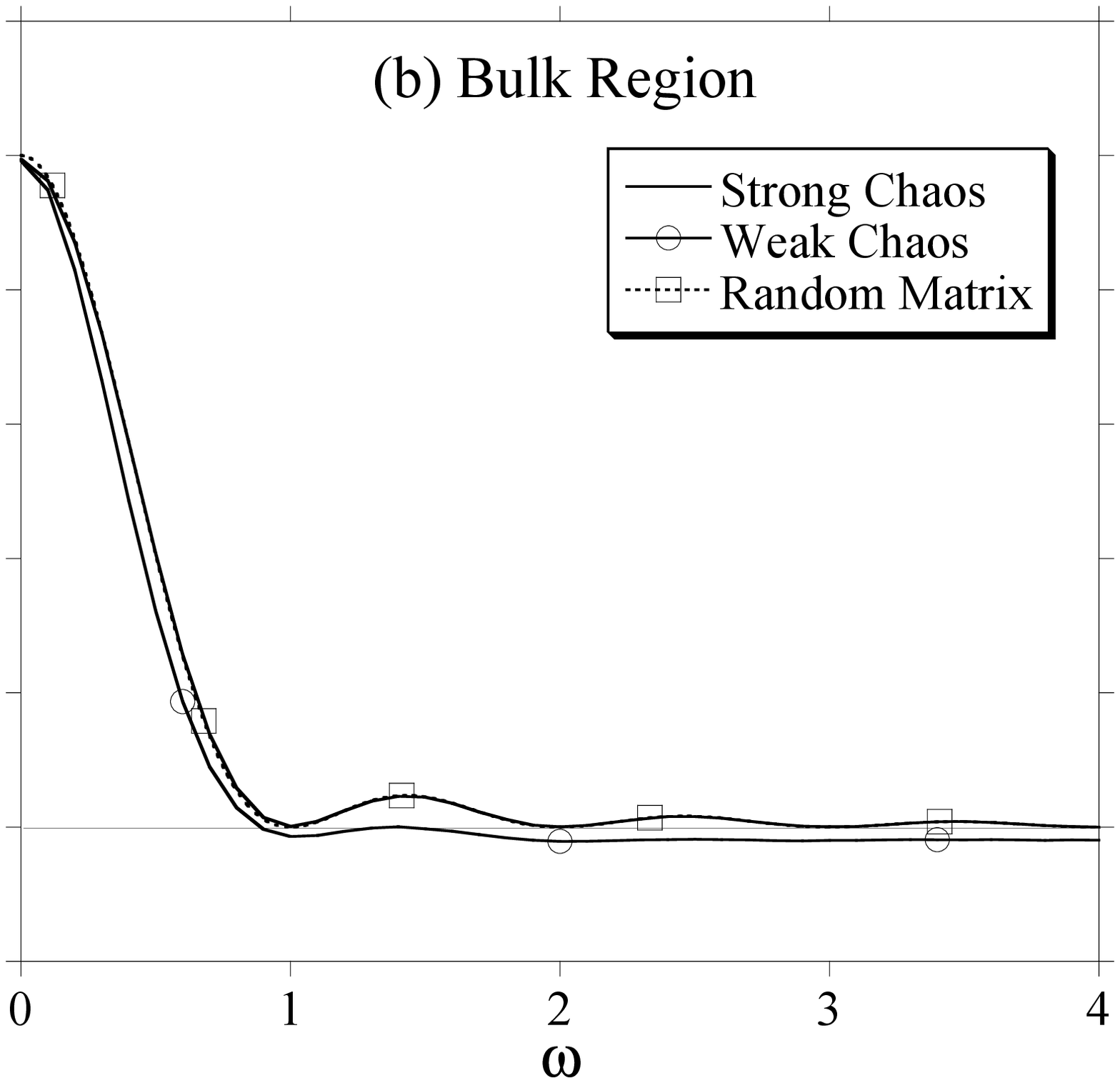}
\includegraphics[height=5.0cm,clip]{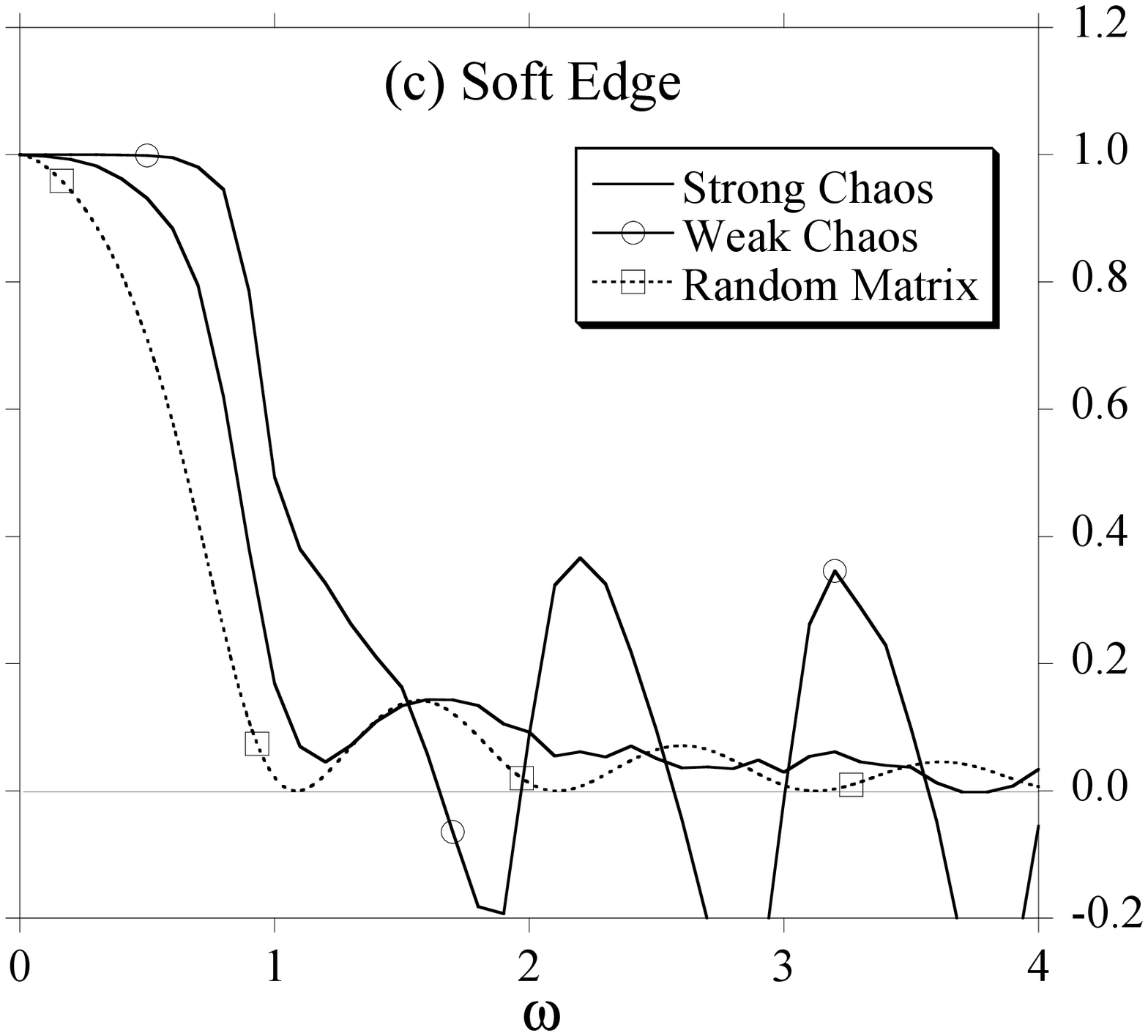}
\caption{\label{fig:TwoCorrelation}
Renormalized two-level cluster function 
$\bar{T}_2(\omega_1,\omega_2)$.
(a) Hard edge: $\bar{T}_2(0,\omega)$ is shown.
(b) Bulk region: the average of $\bar{T}_2(\omega',\omega'+\omega)$ 
over the region $10\le \omega' \le N-10$ is shown.
(c) Soft edge: $\bar{T}_2(N-\omega,N)$ is shown.
}
\end{figure*}

Adachi and Iida \cite{Adachi05} compared the one-level density of 
Schmidt eigenvalues in the coupled standard map with the prediction 
of the random matrix theory, and found that they agree well with each other.
We confirmed their results for our system \cite{Kubotani05}.
In order to see higher order statistics, 
we compare, in Fig.~\ref{fig:TwoCorrelation}, 
the renormalized two-level cluster functions $\bar{T}_2(\omega,\omega')$ 
between the Laguerre unitary ensemble and the ensembles produced by the 
time evolution.
For the hard edge, $\bar{T}_2(0,\omega)$ is depicted
(Fig.~\ref{fig:TwoCorrelation}(a)).
For the strong chaos, 
the renormalized two-level cluster function agrees
with that of the random matrix theory,
although the fluctuations are not negligible.
For the weak chaos, 
it behaves similarly to that of the random matrix theory, 
while the deviation from the random matrix theory is remarkable 
in $\omega\lesssim 1$. 
For the bulk region, in Fig.~\ref{fig:TwoCorrelation}(b), 
$\bar{T}_2(\omega',\omega'+\omega)$ is averaged over 
$\omega'$ in the bulk region.
For the strong chaos, the cluster function agrees well 
with that of the random matrix theory.
However, the cluster function for the weak chaos 
takes negative values after it falls.
This indicates that the Schmidt eigenvalues show weaker avoided crossings.
For the soft edge,  $\bar{T}_2(N-\omega,N)$ is depicted
(Fig.~\ref{fig:TwoCorrelation}(c)).
For both the strong and weak chaos, 
the renormalized two-level cluster functions do not agree 
with that of the random matrix theory.
In particular, an oscillation with large amplitude 
appears for the weak chaos. 

\begin{figure}
\includegraphics[height=5.0cm,clip]{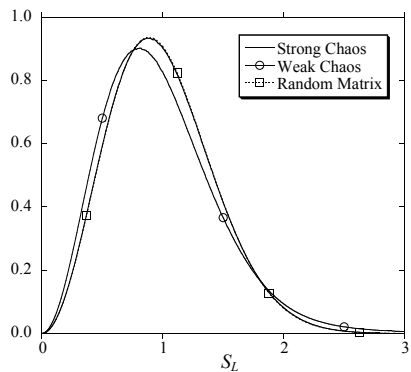}
\caption{\label{fig:LevelSpacing}
Level spacing distribution.
The distributions for the level spacing $S_L$ in the bulk region are 
plotted.
}
\end{figure}

In Fig.~\ref{fig:LevelSpacing}, 
the distribution of level spacing for the bulk region
is shown for the Gaussian unitary ensemble, the ensemble of the strong chaos,
and that of the weak chaos.
Here, we can see that the ensemble of the strong chaos agrees well with 
the Gaussian unitary ensemble.
On the other hand, for the weak chaos, the level spacing distribution 
shifts to smaller level spacings than for the strong chaos. 
This is consistent with the results of the renormalized cluster 
function showing weaker avoided crossings.



We have shown that, when the coupled system exhibits strong quantum chaos, 
the set of the Schmidt eigenvalues approaches 
the Laguerre unitary ensemble.
When the coupled system shows weak quantum chaos,
the set deviates from the the Laguerre unitary ensemble.
This indicates that the Schmidt eigenvalues
for coupled systems exhibiting quantum chaos belong to 
a universality class of the Laguerre unitary ensemble.

The universality of the Schmidt eigenvalues implies that
their distribution will be insensitive to how the coupled system 
is divided into two subsystems interacting with each other.
In other words, the boundary between subsystems has the flexibility that 
the properties of subsystems do not depend significantly
on how we define the boundary.
This is exactly what we expect when the coupled system behaves
in a classically statistical way.
Thus, the universality of the Schmidt eigenvalues can be considered 
as one of the crucial features of the effects called decoherence.

However, the approach to the the Laguerre unitary ensemble
is not complete even for the coupled system exhibiting strong 
quantum chaos.
This is shown by the slight deviation around the soft edge.
This means that dynamics within the Hilbert space spanned 
by these Schmidt eigenvectors still behave in a nonstatistical manner.
Then, the following questions are the next target of our study:
How does the boundary between those behaving statistically 
and those behaving in a nonstatistical way change in time ?
Is there a natural boundary where we can separate 
the coupled system into the two subsystems ?
How does the boundary differ 
depending on the choice of initial conditions ?

There exist a couple of ways to extend the idea of
universality classes into other cases.
One possibility is to consider the coupled system 
where the symmetry exists between the two subsystems.
We expect that the existence of the symmetry will lead to other
universality classes.
Another way of extending our idea is to consider the coupled system
where the sizes of the Hilbert spaces for the subsystems differ 
significantly \cite{Adachi05}. 
This situation corresponds to the cases where the traditional studies of
the statistical physics and the measurement theory consider.
One possibility is that the universality classes can be described
by the extended Laguerre ensemble.
If so, we can think of the following problem:
How does the decoherence effects vary between 
the cases when the system interacts with the other one of the same size 
and those when it interacts with the environment?

The study into the above questions is under progress 
and the results will be published elsewhere.

One of the authors (M. T.) is supported by 
the Japan Society for the Promotion of Science (JSPS)
and Grant-in-Aid for Research on Priority Area 
``Control of Molecules in Intense Laser Fields'' of the Ministry of
Education, Science, Sports and Culture of Japan. 

%



\begin{thebibliography}{99}
\bibitem{EPR}
A. Einstein, B. Podolsky, and N. Rosen, 
Phys. Rev. {\bf 47}, 777 (1935).

\bibitem{Bell}
J. S. Bell, 
Physics {\bf 1}, 195 (1964).

\bibitem{Nielsen}
M. A. Nielsen and I. L. Chuang, 
{\it Quantum Computation and Quantum Information}
(Cambridge Univ. Press, Cambridge, 2000).

\bibitem{Zeh}
E. Joos and H. D. Zeh, Z. Phys. B {\bf 59}, 273 (1985).

\bibitem{Zurek}
W. H. Zurek, Phys. Rev. {\bf D26}, 1862 (1982); 
W. H. Zurek, Phys. Today, {\bf 44}, 36 (1991).

\bibitem{Adachi92}
S. Adachi, in {\it Proceedings of International Symposium on Information 
Physics} (1992), pp.76-83.

\bibitem{Kubotani95}
H. Kubotani, T. Okamura, and M. Sakagami, 
Physica {\bf A214} 560 (1995); 
M. Sakagami, H. Kubotani, and T. Okamura, 
Prog. Theor. Phys. {\bf 95} 703 (1996);
H. Kubotani and M. Den, 
Phys. Lett. {\bf A319} 475 (2003).

\bibitem{Furuya98}
See, e.g., 
K. Furuya, M. C. Nemes, and G. Q. Pellegrino, 
Phys. Rev. Lett. {\bf 80}, 5524 (1998);
P. A. Miller and S. Sarkar, 
Phys. Rev. {\bf E60}, 1542 (1999);
A. Lakshminarayan, 
Phys. Rev. {\bf E64}, 036207 (2001);
A. Tanaka, H. Fujisaki, and T. Miyadera,
Phys. Rev. {\bf E66}, 045201(R) (2002);
H. Fujisaki, T. Miyadera, and A. Tanaka, 
Phys. Rev. {\bf E67}, 066201 (2003);
A. J. Scott and C. M. Caves,
J. Phys. {\bf A36}, 9553 (2003);
S. Bettelli and D. L. Shepelyansky, 
Phys. Rev. {\bf A67}, 054303 (2003);
J. N. Bandyopadhyay and A. Lakshminarayan, 
Phys. Rev. {\bf E69}, 016201 (2004);
P. Jacquod, 
Phys. Rev. Lett. {\bf 92}, 150403 (2004).

\bibitem{Bronk65}
B. Bronk, 
J. Math. Phys. {\bf 5}, 215 (1965).
The joint probability for the Laguerre unitary ensemble 
was first introduced 
for random positive Hermite matrices.

\bibitem{Haake01}
F. Haake, 
{\it Quantum Signatures of Chaos}
(Revised and Enlarged 2nd Edition, Springer-Verlag, Berlin, 2001).

\bibitem{Mehta91}
M. L. Mehta, 
{\it Random Matrices}
(Revised and Enlarged 2nd Edition, Academic Press, New York, 1991).


\bibitem{MatrixComputations}
G. H. Golub and C. F. Van Loan, 
{\it Matrix Computations}
(3rd, Johns Hopkins Univ. Press, Baltimore, 1996).



\bibitem{Kubotani05}
H. Kubotani, M. Toda and M. Adachi (in preparation).

\bibitem{Adachi05}
S. Adachi and S. Iida, in preparation. 
One of the authors (S. A.) and S. Iida have derived 
Eq.~(\ref{LaguerreSigma}) 
without going through the joint distribution in 1990's.

\bibitem{Nagao93}
T. Nagao and K. Slevin, 
J. Math. Phys. {\bf 34}, 2075 (1993).

\end{thebibliography}

\end{document}